\documentstyle[12pt]{article}
\begin{document}
\baselineskip 0.8cm



\begin{center}
{\large\bf  On the $M_T$ scaling of dilepton spectra\\ in high energy
	heavy ion collisions }

\vskip 1.0cm
{\bf D.K. Srivastava\cite{dinesh}, Jicai Pan, V. Emel'yanov\cite{valeri},
 and C. Gale}

\vskip 0.6cm
{\it Physics Department, McGill University \\3600 University St.,
Montr\'eal, QC, H3A 2T8 Canada
}
\end{center}

\vskip 0.6cm
\centerline{\bf Abstract}
\vskip 0.2cm

The so-called $M_T$ scaling for dilepton spectra at RHIC energies is examined.
It is seen
that a proper accounting of the complete set of dilepton producing processes
forces us to abandon the proposed scaling around $M_T$ = 2.6 GeV,
which has been put forward as a possible signature of the presence of
quark-matter. Any substantial transverse expansion in the QGP phase itself
will also offset this scaling behaviour.  The rates of lepton pair production
are time--integrated in Bjorken hydrodynamics and the different sources are
compared against each other.

\newpage
\section{INTRODUCTION}

Quantum Chromodynamics is by now firmly established as the theory of strong
interaction. One of its most spectacular predictions, namely the formation of
quark-gluon plasma (QGP) will be experimentally investigated at the
Relativistic Heavy Ion
Collider under construction at Brookhaven and the proposed
Large Hadron Collider at
CERN. The QGP likely to be produced in these experiments is expected
 to survive
only for a brief duration of at the most a few fm/c, and
a host of suggestions have
been put forward to verify its formation and subsequent evolution.
 The early theoretical descriptions \cite{BJ} have
been tremendously successful in generating activity and ideas. Recently,
approaches have been put forward to treat certain aspects of the
nucleus--nucleus collisions on a microscopic level \cite{bk}. If the QCD
plasma is formed in such collisions and if QCD as a theory admits a first order
phase transition, the system might further pass through a phase mixture
of quarks, gluons and
hadrons \cite{Laszlo}.  Subsequently, the hadrons lose thermal contact and
free-stream towards the detectors.

The space-time evolution of this matter could be
reflected by the yield of high energy photons and dileptons.
These electromagnetic probes are considered reliable because their production
rates are very strongly increasing functions of the temperature
(local energy density) and they escape from the system without
attenuation after they are produced.

These endearing aspects of photons and dileptons have led to a thorough
examination of the processes affecting their production. Thus, for example,
a study of photon--yielding processes in hadrons and QGP led to the
surprising finding that the rate of emission of photons at a given temperature
is phase independent \cite{us}. This necessitated much higher
initial temperatures for the QGP phase for cleaner signals to
emerge \cite{Jan}. Such high
initial temperatures may perhaps be achieved at the RHIC or the LHC
\cite{bk}.

For a long time the dileptons having  large invariant masses were believed
to have their origin in the process $q \bar{q}\rightarrow l\bar{l}$ for the
plasma and in the process $\pi^+\pi^-\rightarrow l\bar{l}$ for the hadronic
matter. As the pion-annihilation process thus envisaged is dominated by the
$\rho$ meson, it immediately led to the suggestion that dileptons having their
invariant masses larger than that for the $\rho$ meson would have their origin
predominantly in the quark matter \cite{S78}.

These early expectations need to be re-examined in the light of
recent findings \cite{Gale}, where for the first time
a complete list of light mesons was used to obtain the rates for dilepton
yield  in a hot hadronic matter. An important consequence
of the quoted work is that the inclusion of vector decays and two-body
reactions
lead to an enhancement of the rate for
dilepton production by a factor of 10--50, compared to the situation when
only pion annihilation is considered, up to an invariant mass $M \approx 3$
GeV.
An immediate outcome of this result is that the safe window for detecting
dileptons from the plasma is now pushed up to a region where Drell--Yan
dileptons are no longer negligible.

These results also force us to abandon the so-called
$M_T$ scaling for dilepton spectra around $M_T =$ 2.6 GeV,
which has been proposed as a signature of the
presence of quark-matter \cite{Ko}. The transverse mass scaling of the lepton
spectrum in high energy heavy ion collisions will occur for an ensemble of
massless quarks in thermal equilibrium, undergoing longitudinal expansion only.
These conditions ensure that the only scale in the problem is the temperature,
$T$. Then $d N / d M^2 d y d^2 q_T$ is a function of $M_T$ only.
However, this feature does not persist
in the hadronic sector where the timelike electromagnetic
form factors explicitly break the scaling law. Nevertheless, there has been
recent hope that the $M_T$ scaling in the pure QCD sector could still be
observed in the net signal \cite{Ko}. As we shall see, this possibility hinges
on the dilepton yielding processes in the nonperturbative regime being
approximated by $\pi^+\pi^-\rightarrow l\bar{l}$, which has
a negligible contribution beyond $M \geq$ 1 GeV. Unfortunately,
the $M_T$ scaling is not observed even for a purely longitudinal
expansion of the system once a full set of dilepton producing rates in
hadronic matter are accounted for. Also, we mention the unavoidable
complications resulting from transverse expansion \cite{trans}. Even in the
absence of the arguments made in this paper, this would seriously challenge the
extraction of a clean signal.

In this paper, we consider the impact of a larger set of
lepton pair emitting processes on lepton spectra and on the hypothesis of $M_T$
scaling.

\section{FORMULATION}

 The dilepton yield calculations for collisions
involving two nuclei have routinely been carried on in the following manner.
  First, the Drell-Yan contribution is either calculated or estimated
by scaling the $pp$  spectrum by an appropriate power of $A$.
 A second contribution is obtained by
integrating the thermal emission rate over the space-time
history of the collision \cite{S78}.  We assume that a thermalized
quark-gluon plasma is formed at some initial temperature $T_0$ at
initial proper time $\tau_0$.  This plasma is then believed to
expand and cool.  Soon
it  reaches the critical temperature $T_c$ for an assumed first-order
chiral symmetry/deconfinement phase transition.  The system is assumed to
continue
to expand at this fixed temperature, deriving the necessary
energy  from the
latent heat of the transition.  Eventually the quark-gluon plasma
is entirely converted to hadronic matter.  After this, the hadrons
might maintain thermal contact for a while, as the matter continues
to expand and cool.  Finally, one estimates a `freeze-out' temperature
$T_f$ at which the hadrons lose thermal contact with each other.
Then they begin free-streaming towards the detectors.  One adds to
the Drell-Yan contribution the emission over the cooling curve,
from $T_0$ to $T_f$, using equations of state for quark-gluon
plasma and hadronic gas.  This is the way in which the dilepton
mass spectrum was computed in ref. \cite{KKMM}, for example.

\subsection{The rates for dilepton production from the QGP and hot mesons}

Dileptons are normally assumed to be produced via the reaction
$q\bar{q} \rightarrow l^+l^-$
in the plasma phase and $\pi^+\pi^- \rightarrow l^+l^-$ in the hadron
phase. In the QCD sector we shall ignore corrections to the Born term.
Those effects, and their consequence on $M_T$ scaling are worthy of interest.
Thus, at this level of approximation,
the corresponding rate for production of
dileptons having an invariant mass $M$ from
the QGP phase is known to be
given by
\begin{equation}
\frac{dN}{d^4xdM^2}=\frac{\sigma_q(M)}{2\ (2 \pi)^4}M^3TK_1(M/T)
{}~\left[1-\frac{4m_q^2}{M^2}\right]\ ,
\label{eq:1}
\end{equation}
where $m_q$ is the mass of the quarks undergoing annihilation,
\begin{equation}
\sigma_q(M)=\frac{4\pi}{3}\frac{\alpha^2}{M^2}
\left[1+\frac{2m_l^2}{M^2}\right]\left[1-\frac{4m_l^2}{M^2}\right]^{1/2}
F_q(M)
\label{eq:2}
\end{equation}
and,
\begin{equation}
F_q(M)=N_c\ (2s+1)\ \sum_f\ e_f^2 =\frac{24}{3}
\end{equation}
for the QGP consisting of u, d, and s quarks. In the above, $m_l$ is the mass
of the leptons, $\alpha$ is the electromagnetic fine structure constant and
$N_c$ is a color factor.  The quark and lepton masses will
ultimately be set to zero.

The corresponding rate for pionic annihilation is obtained by replacing
$m_q$ by $m_\pi$ in Eq. (\ref{eq:1}) and further taking the pionic
form-factor $F_\pi$
in place of $F_q$ in Eq. (\ref{eq:2}). For the pion electromagnetic form factor
we use a recent
parametrization of Biagini et al. \cite{Biagini}.

We have already indicated that the rates of
dilepton yielding
processes in a hot meson gas have recently
been computed \cite{Gale}
by taking a complete set of light mesons
and including decays and two-body reactions. Because of phase space
arguments, these contributions should be dominant. The light pseudoscalar ($P$)
and vector ($V$) mesons we included consist of $\pi$, $\eta$, $\rho$, $\omega$,
$\eta^\prime$, $\phi$, $K$, and $K^*$.
Below the $\pi a_1\ \rightarrow\ e^+ e^-$ threshold,
we have considered  the decays
$\rho\ \rightarrow\ \pi e^+ e^-$,
${K^*}^\pm\ \rightarrow\ K^\pm e^+ e^-$,
${K^*}^0({\bar {K^*}^0})\ \rightarrow\ K^0({\bar K^0}) e^+ e^-$,
$\omega\ \rightarrow\ \pi^0 e^+ e^-$, $\rho^0\ \rightarrow\ \eta e^+ e^-$,
$\eta^\prime\ \rightarrow\ \rho^0 e^+ e^-$, $\eta^\prime\ \rightarrow\ \omega
e^+ e^-$, $\phi\ \rightarrow\ \eta e^+ e^-$, $\phi\ \rightarrow \eta^\prime
e^+ e^-$, $\phi\ \rightarrow \pi^0 e^+ e^-$. We include  all possible reactions
of the type $P$ + $P$ $\rightarrow$ $e^+ e^-$ and $V$ + $V$ $\rightarrow$ $e^+
e^-$. The $V$ + $P$ initial states
considered are $\omega\ \pi^0$, $\rho\ \pi$,   $\phi\ \pi^0$,
   $\omega\ \eta$, $\phi\ \eta$, $\rho^0\ \eta$, $\omega\ \eta^\prime$,
  $\phi\ \eta^\prime$, $\rho\ \eta^\prime$,
${\bar K^*}K$ and $K^*{\bar K}$. Furthermore, the effect of the
$a_1$ meson has been discussed in connection with real photon emission
\cite{xiong,song2}. We have
added the rates corresponding to the initial states $\pi a_1$ and $a_1
a_1$ into electron--positron pairs. The formalism to calculate all the
Feynman amplitudes above is that of effective chiral Lagrangian models
\cite{Gale,song}. Ref. \cite{Gale} contains a complete analysis
of the relative contribution of the processes we have included. The reader will
also find there a discussion of how our Lagrangian parameters have
been adjusted to reproduce experimental data and of form factor issues.

We have found
that the net
rate obtained by summing all the processes scales as $~TK_1(M/T)~$ to a high
degree of accuracy between M=0.3 GeV, and 3 GeV. This is because of
the fact that the
majority of reactions considered in Ref. \cite{Gale} share the pion
electromagnetic form factor. In order to see this most clearly,
 we have plotted in Fig. 1 the
effective form-factor obtained by summing all the channels
enumerated above as,
\begin{equation}
\frac{dN}{d^4xdM^2}=\frac{\sigma_{\rm eff}(M)}{2\ (2\pi)^4}M^3TK_1(M/T)
{}~\left[1-\frac{4m_\pi^2}{M^2}\right]
\end{equation}
with
\begin{equation}
\sigma_{\rm eff}(M)=\frac{4\pi}{3}\frac{\alpha^2}{M^2}~\left[
1+\frac{2m_l^2}{M^2}\right]~
\left[1-\frac{4m_l^2}{M^2}\right]^{1/2}~F_{\rm eff}(M)
\end{equation}
at temperatures of 100, 150, and 200 MeV.
 The scaling with $TK_1(M/T)$
mentioned above is seen
to be satisfied over a reasonable range of invariant mass.
The deviation seen at very low-masses has its origin in the contributions
$\omega\rightarrow\pi^0 l{\bar l},~ \rho\rightarrow\pi l{\bar l}$, and
$\phi\rightarrow\pi^0 l{\bar l}$, which however are only marginal for
large invariant masses. There is a small breaking of our scaling hypothesis
above $M$ = 2.5 GeV because of the $a_1 a_1\ \rightarrow\ e^+ e^-$ threshold.
We have also shown for
comparison the form-factors for the pionic annihilation
and the annihilation of quarks. It is seen that if only pionic annihilation
processes are included, the quark annihilation processes will dominate the
spectra beyond $M\approx 1.5\ $GeV, by two orders of magnitude. However, a
proper accounting of all the mesons present in the hot hadronic matter
makes the identification of the quark--gluon signal considerably more
difficult; we shall return to this point later.

\subsection{The dilepton distributions}

\subsubsection{Results for transverse-mass distributions}

The simplest space-time evolution dynamics is provided
by the longitudinal hydrodynamic expansion model of Bjorken \cite{BJ}.
However after
the proper time $\approx R/c_s$,
where $R$ is the transverse dimension of the system and
$c_s$ is the speed of sound, the rarefaction wave-front
 from the surface will reach
the centre and the transverse expansion \cite{trans}
 of the system can not be ignored.
 The transverse velocity of the fluid elements would  increase monotonically
if there is no change in equation of state. However, a first order phase
transition, as in the case of quark-gluon plasma, slows down
 the transverse expansion of the system considerably.

We have
considered head-on collision of two gold nuclei, and used the criterion
\begin{equation}
T_0^3\tau_0=\frac{2\pi^4}{45\zeta(3)\pi R^2 4a_Q}\frac{dN}{dy_\pi}
\left[f_0+\frac{1-f_0}{r}\right]
\end{equation}
to relate initial temperature ($T_0$) and the time ($\tau_0$)
 to the pion rapidity density (dN/dy$_\pi$). In the above, $f_0$=0 gives the
fraction of the QGP at the initial time, and we have taken $a_Q=47.5\pi^2/90$
for a system consisting of u, d, and s quarks. The fraction $f_0=0$ when
$T_i< T_c$, $0\leq f_0 \leq 1$ when $T_i=T_c$, and $f_0=1$ when $T_i > T_c$.
Furthermore, $r$, the ratio of degrees of freedom between
the QGP phase and the hadronic phases is 47.5/6.8, which corresponds to
hadronic matter having the complete set of light mesons $\pi,\rho,
\omega,\eta,\eta^{'},\phi,$ K, K$^*$, and $a_1$ \cite{Gale}.
As the immediate purpose of this work is to examine the relevance of $M_T$
scaling, we shall use Bjorken's hydrodynamic expansion model in the following.
Now the volume element in Bjorken hydrodynamics is given by,
\begin{equation}
d^4x=d^2x_T\,dz\,dt=\pi R^2\,dy\,\tau d\tau
\end{equation}
In order to see the so-called scaling behaviour of the $M_T$ spectra,
we first consider a case with $T_0$ =250 MeV and $\tau_0$=1 fm/c.
With those values,
the system is initially in the QGP phase during the proper time from
$\tau_0$ to $\tau_q=
(T_0/T_c)^3\tau_0$, in the mixed-phase during $\tau_q$ to $\tau_h=r\tau_q$, and
in the hadronic phase during $\tau_h$ to $\tau_f=(T_c/T_f)^3\tau_h$, beyond
which it undergoes freeze-out. We have taken the critical and freeze-out
temperatures as 160 MeV and 120 MeV, respectively.

Now the number of dilepton pairs having an invariant mass $M$ and transverse
momentum $p_T$, so that $M_T=\sqrt{p_T^2+M^2}$, can be written as
 (see \cite{KKMM} for comparison),
\begin{eqnarray}
\frac{dN}{dM^2dM_Tdy}&=&\frac{\sigma_q(M)M^2M_T}{4(2\pi)^4}
\left[1-\frac{4m_q^2}{M^2}\right]\pi R^{2}3T_0^6\tau_0^{2}M_T^{-6}
[G(M_T/T_0)-G(M_T/T_c)]\nonumber\\
& &+\frac{\sigma_q(M)M^2M_T}{4(2\pi)^4}\left[1-\frac{4m_q^2}{M^2}\right]~
\frac{1}{2}(r-1)\tau_q^{2} \pi R^{2}K_0(M_T/T_c)\nonumber\\
& &+\frac{\sigma_{\rm eff}(M)M^2M_T}{4(2\pi)^4}\left[1-\frac{4m_\pi^2}{M^2}
\right]
\frac{1}{2}r(r-1)\tau_q^{2} \pi R^{2} K_0(M_T/T_c)\nonumber\\
& &+\frac{\sigma_{\rm eff}(M)M^2M_T}{4(2\pi)^4}\left[1-\frac{4m_\pi^2}{M^2}
\right] \pi R^{2}3T_c^{6}\tau_h^2M_T^{-6}[G(M_T/T_c)-G(M_T/T_f)]\ ,\nonumber\\
\end{eqnarray}
where
\begin{equation}
G(z)=z^{3}(8+z^2)K_3(z)~~.
\end{equation}

In Fig. 2 we have plotted the results for the transverse mass distribution
of dileptons for a fixed $M_T$ = 2.6 GeV \cite{Ko}, against $p_T$.
Note that this way of plotting is convenient: if the scaling holds this
plot will be a straight horizontal line. We show results obtained
when only the pionic annihilation process is included for the hadronic
matter, and when the entire spectrum of hadronic processes leading to
dileptons are considered. If we restrict our analysis to the pion--pion
process, the total contribution obtained by summing up the quark matter and
hadronic matter signals will exhibit scaling.
However, taking account of all the processes in the hadronic matter leads to
a considerable enhancement of dilepton production even up to an invariant mass
of 2 GeV (lower values of $p_T$ on this figure) \cite{Gale}. No scaling of the
$M_T$
spectra will be seen as the QGP contribution, which itself scales with $p_T$
for
the case of no transverse expansion, is only
a small fraction of the total contribution. In this sense, the apparent
flatness below $p_T$ = 1.5 GeV/c is accidental and not a reflection of the
QCD $M_T$ scaling. The bulk of the signal is from the
hadronic sector, where the hadronic electromagnetic
form factors break the scaling. It
should also be remembered that, even if
the lifetime of the QGP is large, as it would be in the case of a higher
initial temperature, the transverse velocity of the
fluid could then be substantial even during the QGP phase which would
furthermore
offset the $M_T$ scaling property even for the QGP contribution.

Thus we conclude that the $M_T$ scaling of the dilepton spectra,
which has been proposed as a signature of the presence of quark-matter, does
not
hold at the temperatures considered in this work, once a more complete
description of dilepton--yielding processes is achieved. We do not display in
this work the
$M_T$ feature of the Drell--Yan dileptons. We shall return to this point in the
discussion.

\subsubsection{The invariant mass distribution}

The invariant mass distribution of dileptons provides a very
useful parametrization of the data, as they are relatively less affected by
the models for space-time evolution. The final results are easily written
for the initial state in QGP as,
\begin{eqnarray}
\frac{dN}{dM^2dy}&=& \frac{\sigma_q(M)}{2(2\pi)^4}M^3\left[1-\frac{4m_q^2}{M^2}
\right] \pi R^{2} 3T_0^6\tau_0^2M^{-5}[H(M/T_0)-H(M/T_c)]\nonumber\\
& &+\frac{\sigma_q(M)}{2(2\pi)^4}M^3\left[1-\frac{4m_q^2}{M^2}\right]
\pi R^{2} \frac{1}{2}(r-1)\tau_0^{2}T_cK_1(M/T_c)\nonumber\\
& & +\frac{\sigma_{\rm
eff}(M)}{2(2\pi)^4}M^3\left[1-\frac{4m_\pi^2}{M^2}\right]
\pi R^{2}\frac{1}{2}r(r-1)\tau_0^{2}T_cK_1(M/T_c)\nonumber\\
& & +\frac{\sigma_{\rm
eff}(M)}{2(2\pi)^4}M^3\left[1-\frac{4m_\pi^2}{M^2}\right]
\pi R^{2}3T_c^6\tau_c^2M^{-5}[H(M/T_c)-H(M/T_f)]\ ,\nonumber\\
\end{eqnarray}
where
\begin{equation}
H(z)=z^2(8+z^2)K_0(z)+4z(4+z^2)K_1(z)\ .
\end{equation}

We plot in Fig. 3 the mass spectrum for the same $\{ \tau_0 , T_0
\}$ initial conditions as in the previous section.
We have also shown the Drell-Yan spectrum
for collision of two lead nuclei at $\sqrt{s}$ = 200A GeV, with structure
functions obtained from the set I of Duke and Owens\cite{DO}. It is clear
from this figure
that our thermal QCD signal has little or no chance of shining through the
hadron gas and Drell--Yan ``backgrounds''. Then, for a somewhat extreme
viewpoint, we have also plotted the invariant mass spectrum for
dileptons, taking initial
temperature $T_0$ as 500 MeV and the initial time $\tau_0=1/3T_0$, appropriate
for energies reached at RHIC for collisions involving two gold nuclei
\cite{dinesh2}. Looking at Fig. 4, we immediately realize
that the invariant mass window for
seeing the emissions
from the QGP are pushed to beyond 2 GeV, once the full spectrum of dilepton
yielding reactions is included.
We note the
discomforting feature that, even at such temperatures,  the emissions from
the QGP are neither much larger
than the Drell-Yan contribution, nor are they significantly larger than the
emissions from the hadronic matter. This demands a very careful
analysis of the dilepton spectra before it can be used as a signature of
QGP at RHIC energies. We do not plot the $M_T$ scaling features of our solution
with this high initial temperature: there, transverse expansion effects could
certainly not be ignored. The transverse expansion influence on
the net invariant mass spectrum also needs to be computed, as it will shorten
the lifetime of the hadronic and mixed phases \cite{ruus}.

\section{Discussion}

We have addressed in this paper the issue of the proposed $M_T$ scaling of the
dilepton spectrum in high energy heavy ion collisions. We have found that
the integrated lepton pair distributions would not keep any memory of the $M_T$
scaling properties of the original plasma phase. This is because of the
sheer size of the hadronic gas signal, and because the
scaling hypothesis is
violated strongly in the nonperturbative sector, owing to the structure of the
timelike electromagnetic form factors. Of course, any calculation such as ours
is an approximate one. However, we believe that the approximations we have made
would have made the scaling easier to observe.

Firstly, we have not attempted a quantitative calculation of transverse flow
stemming from hydrodynamic effects. To reiterate, any flow features will spoil
the scaling properties. If the quark--gluon signal is helped by dialing
a high initial temperature in our calculations, this indicates (i) a somewhat
longer lived QCD phase where flow will develop on its own and, (ii) the
importance of the relaxation mechanisms. This brings us to our second point.
In our transverse mass
spectra we have neglected the Drell--Yan signal. It is easy to verify using
simple expressions that the Drell--Yan signal violates $M_T$ scaling. This
is easily understood, as one of the premises for scaling is thermal
equilibrium.
However, the simple Drell--Yan calculations might not be so
relevant for the high energy
collisions of heavy nuclei if the original parton distribution is modified
during a nucleus--nucleus collisions.  Such effects could surely modify the
Drell--Yan estimates. Perhaps an adequate attempt to consider parton
rescattering effects is provided by the Parton Cascade Model
(PCM) \cite{bk}. In
this picture, the partons start in a completely nonequilibrium state and they
are subsequently driven towards relaxation by  continuous processes. In the
PCM, the $M_T$ scaling features have been analyzed \cite{gk2}. The result there
is that the scaling
hypothesis is also strongly violated, even at the parton level. This state of
affairs may owe to several facts. Among others, they are the lack of thermal
and chemical
equilibration, and the existence of additional scales in the problem, provided
by infrared cutoff parameters and higher order QCD corrections. Also, processes
that do not scale may also contribute \cite{asak}. Combining the
above considerations
with our own results,   we are forced to the somewhat pessimistic conclusion
that
there is little hope that the observed lepton pair spectrum could show any
evidence of ``genuine'' scaling. Conversely, any flatness in the transverse
mass spectrum, plotted against transverse momentum, could not be interpreted
unambiguously as a relic of a deconfined QCD phase.

\section*{Acknowledgements}

We are happy to acknowledge useful communications with H. Eggers, K. Geiger, K.
Haglin, C.M. Ko, P. V. Ruuskanen and C. Song. This work was
supported in part by the
Natural Sciences and Engineering
Research Council of Canada, in part by the FCAR fund of the Qu\'ebec
Government, and in part by a NATO Scientific Exchange Award.

\newpage



\newpage
\leftline{\large\bf Figure captions}
\begin{description}

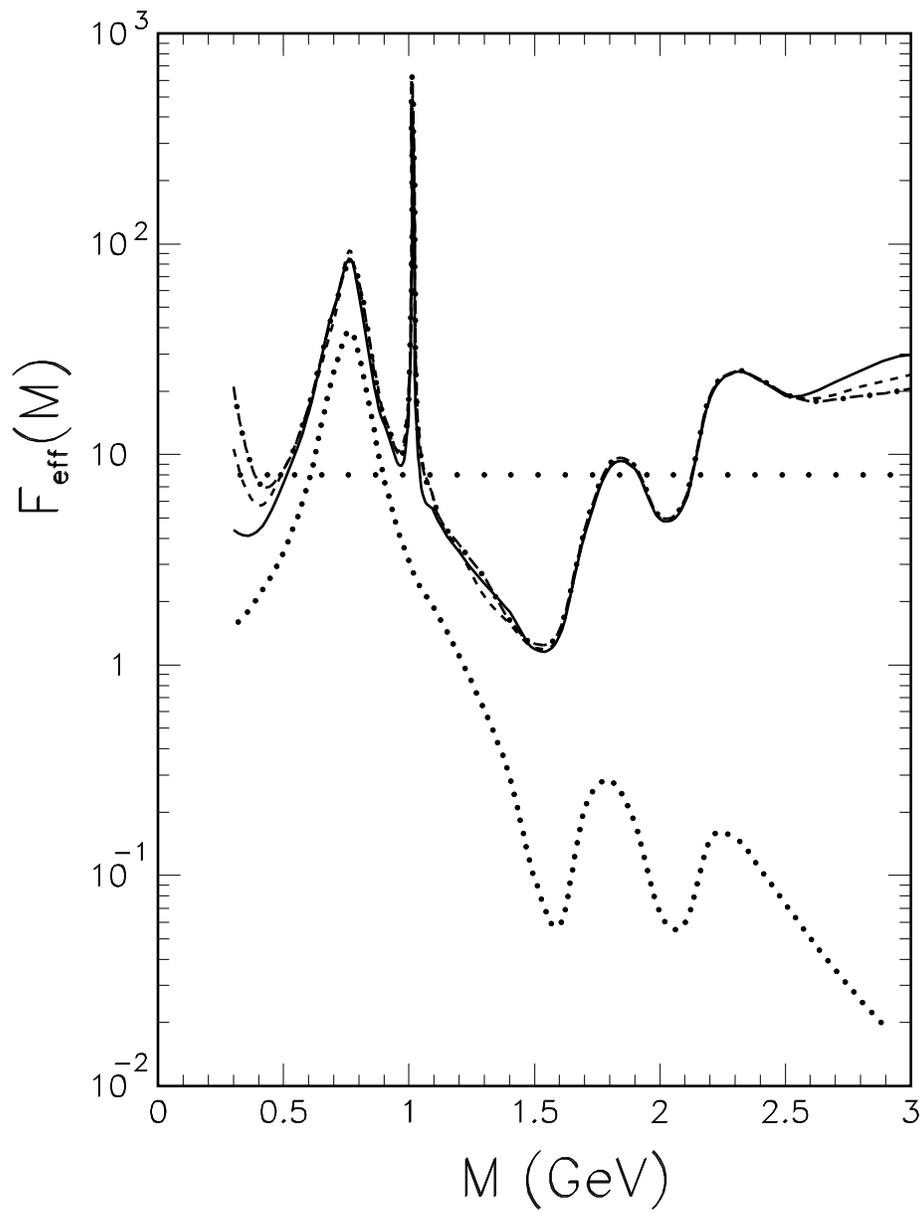
\item[Fig. 1]
The effective form-factor for dilepton yielding
processes in hadronic matter when the rates obtained in Ref. [7] are
written as Eq. (2.4). The dashed--dotted, dashed and solid curves represent the
effective form--factor for $T$= 150, 200 and 250 MeV, respectively. The
tight--dotted curve is the form-factor for pionic annihilation
from Biagini et al. [11]. The loose--dotted curve represents the
quark--antiquark
``form factor''.

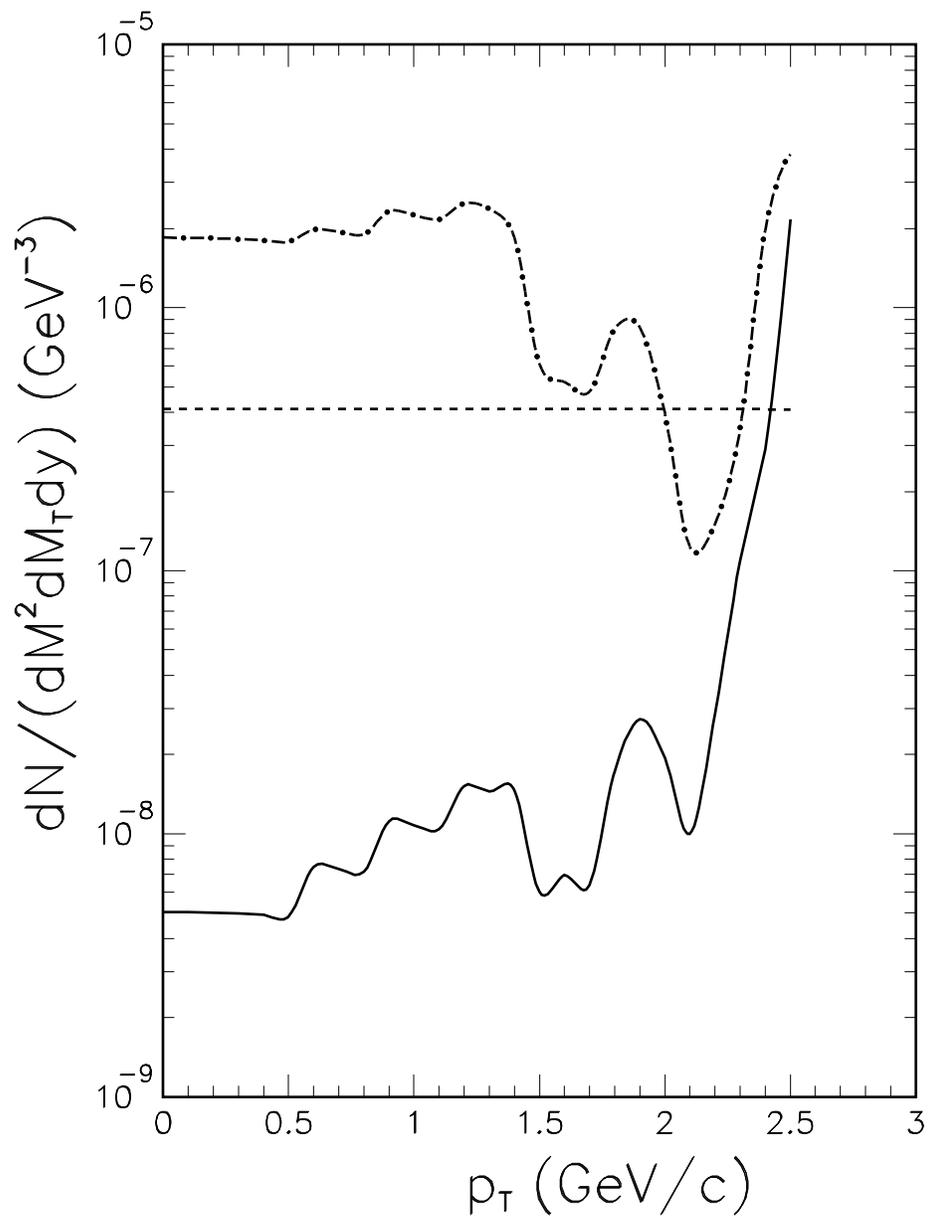
\item[Fig. 2]
Transverse-mass spectra for  dileptons, for
contributions from quark--annihilation (horizontal dashed line),
from the full set of
mesons (dashed--dotted curve), and from the pion gas approximation (solid
curve). The temperature is 250 MeV and $\tau_0$ = 1 fm/c. The transverse
mass is fixed at $M_T$ = 2.6 GeV [8].

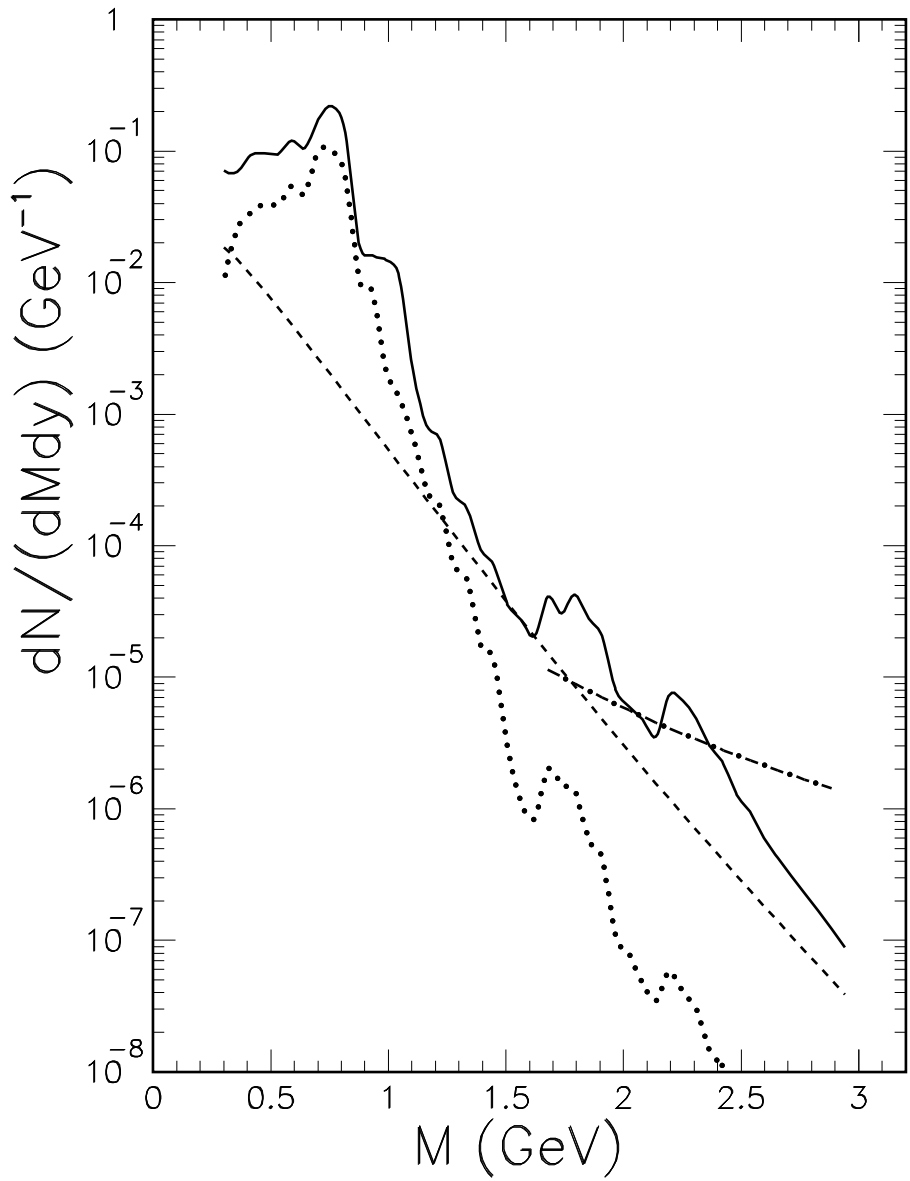
\item[Fig. 3]
Dilepton mass spectra. We show the contributions
from quark-matter (dashed line), pion annihilation (dotted curve), Drell-Yan
(dashed--dotted curve) processes and from all the
hadronic processes considered in Ref. [7] (solid curve). The temperature and
formation time are the same as in the previous figure.

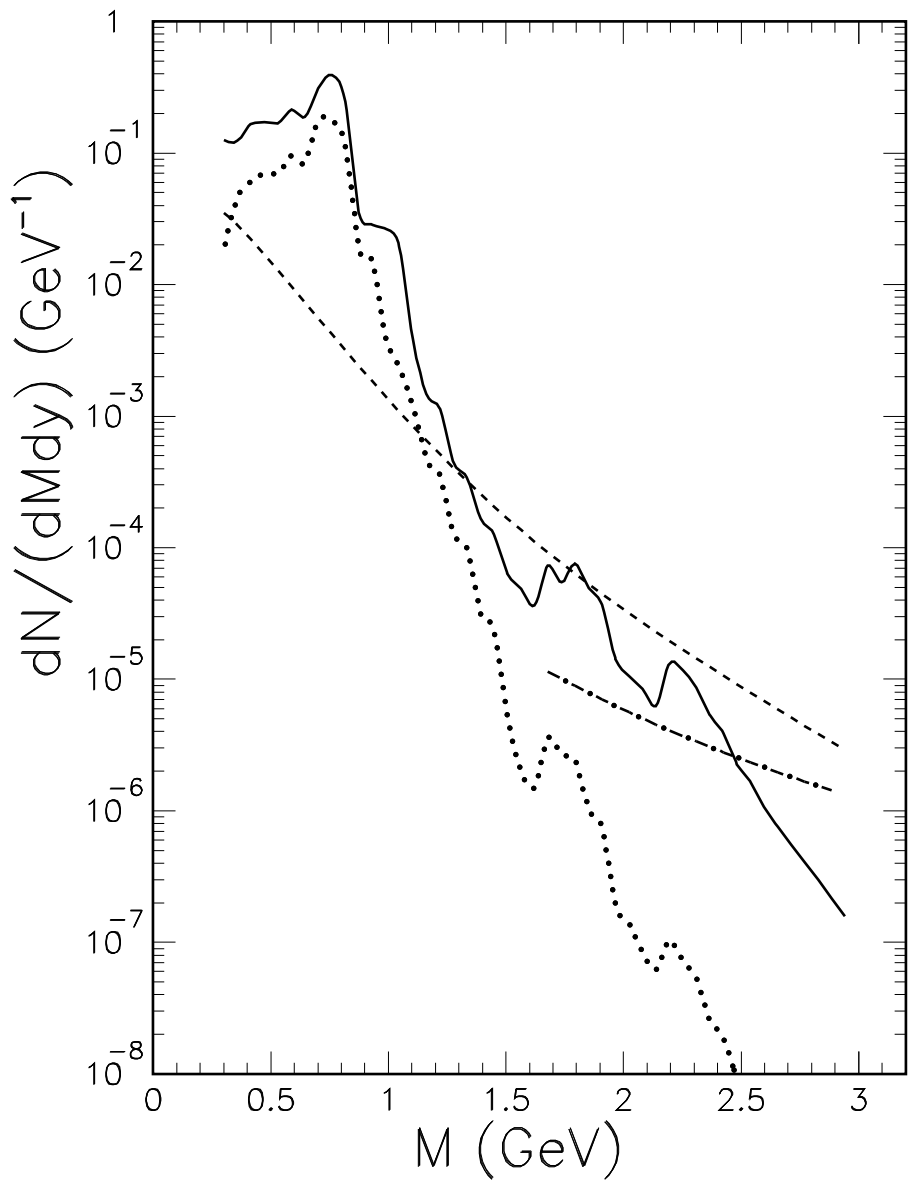
\item[Fig. 4]
Same caption as in Fig. 3, but for $T_0$ = 500 MeV and
$\tau_0$ = 1/3$T_0$.

\end{description}

\end{document}